\documentclass[a4paper,fleqn,usenatbib]{mnras}

\usepackage{amsmath}
\usepackage{mathptmx}
\usepackage{txfonts}
\usepackage[T1]{fontenc}
\usepackage{graphicx}	
\usepackage{amssymb}	
\usepackage[normalem]{ulem}
\usepackage{longtable}
\usepackage{multirow}
\usepackage{lmodern}

\newcommand{\be}{\begin{equation}}
\newcommand{\e}{\end{equation}}
\newcommand{\bear}{\begin{eqnarray}}
\newcommand{\ear}{\end{eqnarray}}

\newcommand{\hmpc}{{\, h^{-1}\, {\rm Mpc}}}

\def\aj{AJ}
\def\apj{ApJ}
\def\apjs{ApJS}
\def\jcap{JCAP}
\def\mnras{MNRAS}
\def\aap{A\&A}

\def\nat{Nature}      
\def\apjs{ApJS}
\def\apjl{ApJ Letters}

\title[Galaxy colour and the cosmic web] {Exploring galaxy colour in
  different environments of the cosmic web with SDSS}

\author[Pandey, B. and Sarkar, S.] { Biswajit
  Pandey$^1$\thanks{biswap@visva-bharati.ac.in} and Suman
  Sarkar$^{1}$\thanks{suman2reach@gmail.com} \\$^1$ Department of
  Physics, Visva-Bharati University, Santiniketan, Birbhum, 731235,
  India }

 \date{\today}

 \pubyear{2020}
  
\begin{document}
\label{firstpage}
\pagerange{\pageref{firstpage}--\pageref{lastpage}}      
\maketitle
       
\begin{abstract}
We analyze a set of volume limited samples from the SDSS to study the
dependence of galaxy colour on different environments of the cosmic
web. We measure the local dimension of galaxies to determine the
geometry of their embedding environments and find that filaments host
a higher fraction of red galaxies than sheets at each luminosity. We
repeat the analysis at a fixed density and recover the same trend
which shows that galaxy colours depend on geometry of environments
besides local density. At a fixed luminosity, the fraction of red
galaxies in filaments and sheets increases with the extent of these
environments. This suggests that the bigger structures have a larger
baryon reservoir favouring higher accretion and larger stellar
mass. We find that the mean colour of the red and blue populations are
systematically higher in the environments with smaller local dimension
and increases monotonically in all the environments with
luminosity. We observe that the bimodal nature of the galaxy colour
distribution persists in all environments and all luminosities, which
suggests that the transformation from blue to red galaxy can occur in
all environments.
\end{abstract}
       

\section{Introduction}

The present Universe is full of galaxies which come in various shape
and size with different mass, luminosity, colour, star formation rate,
metallicity and HI content. Understanding the galaxy properties and
their evolution is an important goal of cosmology. The modern galaxy
surveys (2dFGRS,\citealt{colless}; SDSS, \citealt{strauss02}) reveal
that the galaxies are distributed in the cosmic web \citep{bond96}
which is an interconnected weblike network comprising of different
types of environments such as filaments, sheets, knots and voids. The
galaxy properties vary across the different environments in the cosmic
web. For example, the well known density-morphology relation reveals
that the ellipticals are preferably found inside the dense groups and
clusters whereas the spirals are intermittently distributed in the
fields \citep{hubble, zwicky, Oemler, dress, gotto}. These findings
are further supported by other studies with two-point correlation
function \citep{will, brown, zehavi}, genus statistics \citep{hoyle1,
  park1} and filamentarity \citep{pandey1, pandey2} of the galaxy
distribution. It is now well known that many other galaxy properties
are strongly sensitive to their environment \citep{davis2,
  guzo,zevi,hog1, blan1, einas2, kauffmann, mocine, koyama}. The
formation and evolution of galaxies are known to be driven by
accretion, tidal interaction, merger and various other secular
processes. These physical processes are largely determined by the
environment of the galaxies. The environment thus play a central role
in the formation and evolution of galaxies and the study of the
environmental dependence of the galaxy properties provides crucial
inputs to the theories of galaxy formation and evolution.

A number of studies have been carried out to understand the
environmental dependence of galaxy colour. The galaxy colour
distribution is well fit with a double Gaussian distribution
\citep{balogh2004} which can be used to divide the galaxies into red
and blue populations. It has been shown that the blue galaxies reside
preferentially in low-density regions whereas the red galaxies inhabit
the high-density regions \citep{hogg2004, baldry2004, blan2,
  ball2008}. \citet{parkc} find that the galaxy colour is not
sensitive to local density at fixed luminosity and
morphology. \citet{balogh2004} show that when the luminosity is fixed,
the fraction of galaxies in the red population is a strong function of
local density. \citet{cooper} find a highly significant correlation
between stellar age and environment at fixed stellar mass for the red
galaxies. \citet{bamford} find that the galaxy colour is highly
sensitive to environment at a fixed stellar mass. They reported that
irrespective of morphology, the high stellar mass galaxies are mostly
red in all environments whereas the low stellar mass galaxies are
mostly blue in low density environment and red in high density
environment.

Most of these studies use the local density as a proxy for the
environment. Different methods are often used to define the
environment based on the scale being probed \citep{muldrew}. The
various environments of the cosmic web are characterized by the
density as well as their geometry. The clusters represent the densest
regions in the cosmic web followed by the filaments, sheets and
voids. The galaxy clusters are the dense knots which reside at the
intersection of elongated filaments. The filaments in the cosmic web
are located at the intersection of sheets which encompass large empty
regions or voids. Studies with N-body simulations \citep{arag10,
  cautun14, ramachandra} suggest that matter successively flows from
voids to walls, walls to filaments and then channelled along the
filaments onto the clusters. The dark matter halos formed at various
environments of the cosmic web are known to have different mass, shape
and spin due to the influence of their large-scale environment
\citep{hahn1}. The galaxies are assumed to form at the centre of these
dark matter halos via cooling and condensation of baryons
\citep{white}. In the halo model, the mass of the dark matter halo is
believed to be the single most important parameter which determines
the properties of a galaxy \citep{cooray}.  However the clustering of
the dark matter halos depend on their assembly history \citep{croton,
  gao, musso, vakili} besides their mass. This implies that the
environmental dependence of the galaxy properties may extend beyond
the local density and the large-scale environments in the cosmic web
may impart significant influence on the formation and evolution of
galaxies. However there are no universal measure for characterizing
the large-scale environments in the cosmic web. Some of the existing
statistical tools designed for this purpose are the Shapefinders
\citep{Sahni1998}, the statistics of maxima and saddle points
\citep{colombi}, the multiscale morphology filter \citep{arag}, the
skeleton formalism \citep{novikov} and the local dimension
\citep{sarkar}.

In the present work, we use the local dimension \citep{sarkar} to
quantify the different environments in the cosmic web. A recent
analysis with local dimension find that the sheets are the most
prevalent pattern in the SDSS galaxy distribution which can extend
up to $90 \hmpc$ \citep{sarkar19}. They also show that the straight
filaments extend only up to a length scale of $30 \hmpc$. The different
structural components of the cosmic web differ in density, morphology
and size. Each of these components provides an unique environment for
galaxy formation and evolution. The role of the large-scale
environment in this context is not yet settled. \citet{lupa} find that
the properties of the brightest group galaxies in the SDSS depend on
their embedding large-scale structures. Using SDSS, \citet {scudder}
find that the star formation rates in isolated groups and the groups
embedded in superstructures are significantly different. \citet{yan}
analyze the SDSS data to find that the tidal environment of large
scale structures do not influence the galaxy
properties. \citet{eardley} studied the luminosity
  function in different environments using GAMA survey and find no
  evidence of the influence of environments beyond local density.
Recently, \citet{lee} show that the elliptical galaxies in the
sheetlike environment dwell in the regions with the highest tidal
coherence. \citet{pandey17} analyze the Galaxy Zoo \citep{lintott}
data using information theoretic measures to find that morphology and
environment exhibit a synergic interaction up to a length scales of $
\sim 30 \hmpc$.

The Sloan Digital Sky Survey (SDSS) is the largest spectroscopic and
photometric galaxy redshift survey to date. It provides an
unprecedented view of the cosmic web by precisely mapping millions of
galaxies in the nearby Universe. In the present work, we analyze the
spectroscopic data from the SDSS DR16 \citep{ahumada}.  The primary
aim of the present work is to identify the galaxies in different
environments of the cosmic web using local dimension and explore the
variations of red and blue galaxy fractions and their average colours
across these environments. This would reveal how the galaxy colour is
influenced by the different environments in the cosmic web. We study
how the red and blue fractions vary in different geometric
environments. We carry out the same analysis at a fixed local density
to see if geometry of environments have any role in deciding the
colours of galaxies. We also study the variations of red and blue
fractions across different environments characterized by their local
denisty. Besides density and geometry, the size of the different
structural components of the cosmic web may also play a role in
shaping the colour of a galaxy. We address this by measuring the red
and blue fractions across different environments with increasing
length scales keeping the luminosity of the galaxies fixed.  We also
study the bimodal nature of the colour distribution across different
environments at different luminosities and examine how the mean and
dispersion of galaxy colour are sensitive to these parameters. The
analysis presented in this work would help us to understand the
environmental factors besides the local density which may influence
the colour of galaxies.

We use a $\Lambda$CDM cosmological model with $\Omega_{m0}=0.305$,
$\Omega_{\Lambda0}=0.695$ and $h=0.674$ to calculate distances from
redshifts throughout the analysis.

A brief outline of the paper is as follows. The method of analysis and
the data are described in Section 2 and Section 3 respectively. The
results of the analysis are discussed in Section 4 and we present our
conclusions in Section 5.

\begin{figure*}
\resizebox{14 cm}{!}{\rotatebox{0}{\includegraphics{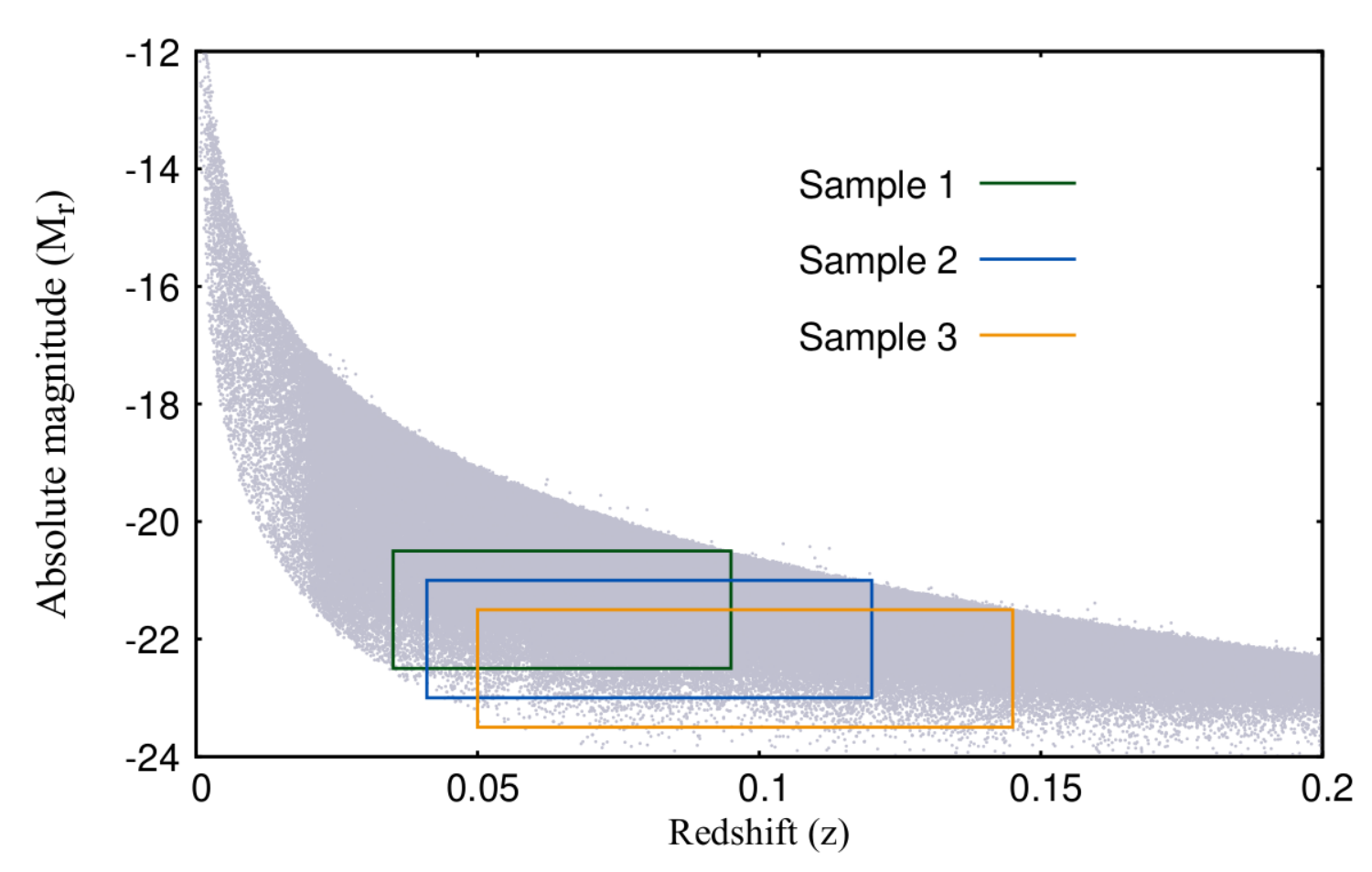}}}
\caption{This figure shows the definition of the volume limited
  samples (\autoref{tab:samples}) in the redshift-absolute magnitude
  plane.}
\label{fig:zM}
\end{figure*}

\section{METHOD OF ANALYSIS}

We use the local dimension to isolate the galaxies residing in
different geometric environments of the cosmic web. We count the
number of galaxies $N(<R)$ within a sphere of radius $R$ centered on
each galaxy. The radius of the sphere is varied within a specific
range $R_1\leq R \leq R_2$ and the number counts $N(<R)$ are measured
for a number of different radius $R$ at equally spaced interval of
$0.5 \hmpc$ within this range. Only the galaxies for which all the
spheres with radius in this range remain completely inside the survey
boundary are treated as valid centres. For any volume limited galaxy
sample this would be possible for only a subset of galaxies and the
number of classifiable galaxies would decrease with increasing
$R_2$. In the present analysis we choose $R_1=5 \hmpc$ and then
gradually increase $R_2$ in steps of $5 \hmpc$ up to the radius of the
largest sphere that would fit inside the survey region.

The galaxies are embedded in different geometric environments in the
cosmic web and the number count $N(<R)$ around a galaxy is expected to
scale as,
\begin{eqnarray}
 N(<R)= A R^{D}
\label{eq:ld1}
\end{eqnarray}
where A is a constant and the exponent $D$ is the local dimension
\citep{sarkar}. The local dimension $D$ tells us about the geometry of
the embedding environment within a length scale range $R_1 \hmpc \leq
R \leq R_2 \hmpc$. We consider only the centres for which there is at
least 10 galaxies within this radius range. For each centres, We
obtain the best fit values of A and $D$ using the least-square
fitting. We calculate the associated $\chi^2$ values using the
observed and the fitted values of $N(<R)$. Only the galaxies for which
the chi-square per degree of freedom $\frac{\chi^2}{\nu} \leq 0.5$ are
considered in the present analysis. Ideally one would expect $D=1$ for
galaxies residing in filaments, $D=2$ for the galaxies residing in
sheets and $D=3$ for the galaxies residing in a three dimensional
volume. However the size and shape of these structures vary
considerably within the cosmic web. Besides there would be galaxies in
the intermediate environments (e.g. junction of a sheet and a
filament). In this case the measuring sphere would incorporate
multiple type of structures within it. We classify the environment of
a galaxy in five different classes (\autoref{tab:gclass}) depending on
their local dimension. We assign a specific range of local dimension
to each class. The galaxies in class $D1$ are residing in
filaments. These galaxies will primarily represent the middle parts of
the straight filaments, which are not closer to the nodes. The
$D1$-type galaxies may also represent short tendrils of galaxies which
are found to link filaments together and penetrate into voids as
coherrent structures \citep{alpaslan14}. The $D2$ type galaxies are
embedded in sheets. The $D3$ type galaxies are mostly field galaxies
in the low density regions. The $D1.5$ type galaxies have intermediate
local dimension between filaments and sheets whereas $D2.5$ galaxies
have local dimension in between sheets and fields.  The environment of
a galaxy as indicated by its local dimension would depend on the
length scales under consideration.

We isolate the galaxies in different environments and segregate them
into red and blue populations in each environment based on their $u-r$
colour. We use an optimal colour cut $(u-r)=2.22$ prescribed by
\citet{strateva} to classify the red and blue galaxies. The galaxies
with $u-r < 2.22$ are identified as blue whereas the ones with $u-r
\geq 2.22$ are termed as red. We use the observed colour for
simplicity keeping in mind the similar redshift range of the volume
limited samples used in this work.

Red galaxies are known to reside in relatively denser environments. We
also consider the effects of local density on the fraction of red and
blue galaxies. We would like to test if the effects of environment
extend beyond the local density and whether morphology of large scale
structures play any role in deciding the colours of galaxies.

The local density at each galaxy position is
  estimated by using $k^{th}$ nearest neighbour method
  \citep{casertano85}. In this method, the local number density is
  defined as,
\begin{eqnarray}
n_k = \frac{k-1}{V(r_k)}  
\label{eq:knn}
\end{eqnarray} 
where the $k^{th}$ nearest neighbour distance to the galaxy is $r_k$
and $V(r_k)=\frac{4}{3}\pi r_k^3$. We choose $k=5$ for the present
analysis. We classify the galaxies into five different categories
based on the local density at their locations. The criteria for each
density class are provided in \autoref{tab:nnden}.

We calculate the fraction of red and blue galaxies in each geometric
environment on different length scales for each of the volume limited
samples described in the next section. We prepare 10 jackknife samples
from each volume limited sample to estimate the errorbars for our
measurements. Each jackknife sample is prepared by randomly deleting
$25\%$ galaxies from the original sample.

\begin{table*}
\caption{This table shows the classification of the five types of
  galaxy environment based on their local dimension.}
\label{tab:gclass}
\begin{tabular}{|l|c|c|c|c|c|}
\hline
Local dimension : & $ 0.75 \le D < 1.25 $ & $1.25 \le D < 1.75$  &  $ 1.75 \le D < 2.25$  &  $ 2.25 \le D < 2.75 $   & $ D \ge 2.75 $ \\
Classified as: & $D1$ & $D1.5$ & $D2$ & $D2.5$ & $D3$ \\
\hline
\end{tabular}
\end{table*}

\begin{table*}{}
\caption{This table summarizes the three volume limited samples extracted from SDSS DR16.}
\label{tab:samples}
\begin{tabular}{ccccccc}
\hline
Volume	&	Absolute  	&  					& Number of	& 		Number 				& Mean inter-galactic	& Radial\\
limited	& 	magnitude 	& 	Redshift ($z$)	& galaxies  	& 		density				& separation 	& size\\
samples	&	($M_r$) 		&     				& 	& 	( $ h^{3} {\rm Mpc} ^{-3} $ ) 	& ($\hmpc$) 		& ($\hmpc$)\\
\hline
 
Sample 1 &$-20.5 \ge M_r \ge -22.5$ & $0.035 \le z \le 0.095$ & $ 90270$ & $9.72 \times 10^{-3}$ & $4.69$ & $174.40$ \\
Sample 2 &$-21   \ge M_r \ge -23  $ & $0.041 \le z \le 0.120$ & $104137$ & $5.61 \times 10^{-3}$ & $5.63$ & $227.93$ \\
Sample 3 &$-21.5 \ge M_r \ge -23.5$ & $0.050 \le z \le 0.145$ & $ 92848$ & $2.90 \times 10^{-3}$ & $7.01$ & $271.47$ \\
\hline
\end{tabular}
\end{table*}

\begin{table*}{}
\caption{This table summarizes the five bins used for classifying the
  galaxies based on their local number density.}
\label{tab:nnden}
\begin{tabular}{cccccc}
\hline
Local density class & $\eta_1$ & $\eta_2$ &$\eta_3$ & $\eta_4$ & $\eta_5$ \\
\hline
Local density range ( in $h^3 Mpc^{-3}$ ) & $ n_{5} < 0.03 $  & $0.03 \le n_{5} < 0.06 $  & $0.06 \le n_{5} < 0.09$  & $0.09 \le n_{5} < 0.12$ & $ n_{5} \geq 0.12$\\
\hline
\end{tabular}
\end{table*} 

\begin{figure*}
\resizebox{18 cm}{!}{\rotatebox{0}{\includegraphics{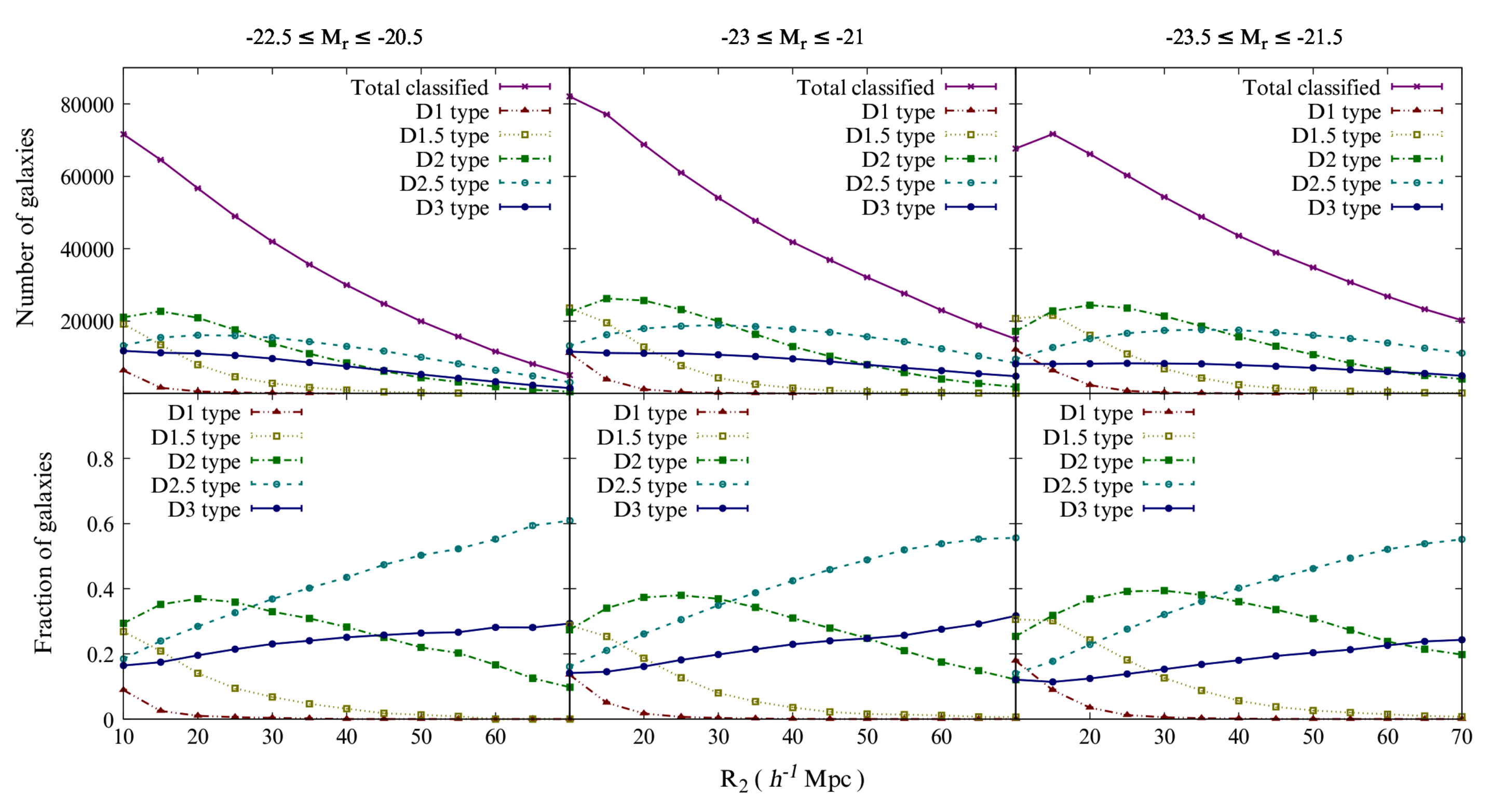}}}
\caption{The top three and the bottom three panels of this figure
  respectively show the variations in the number and fraction of
  classified galaxies belonging to different types of environment
  (\autoref{tab:gclass}) as a function of $R_2$ for the three volume
  limited samples (\autoref{tab:samples}).}
\label{fig:frac}
\end{figure*} 

\begin{figure*}
\resizebox{18 cm}{!}{\rotatebox{0}{\includegraphics{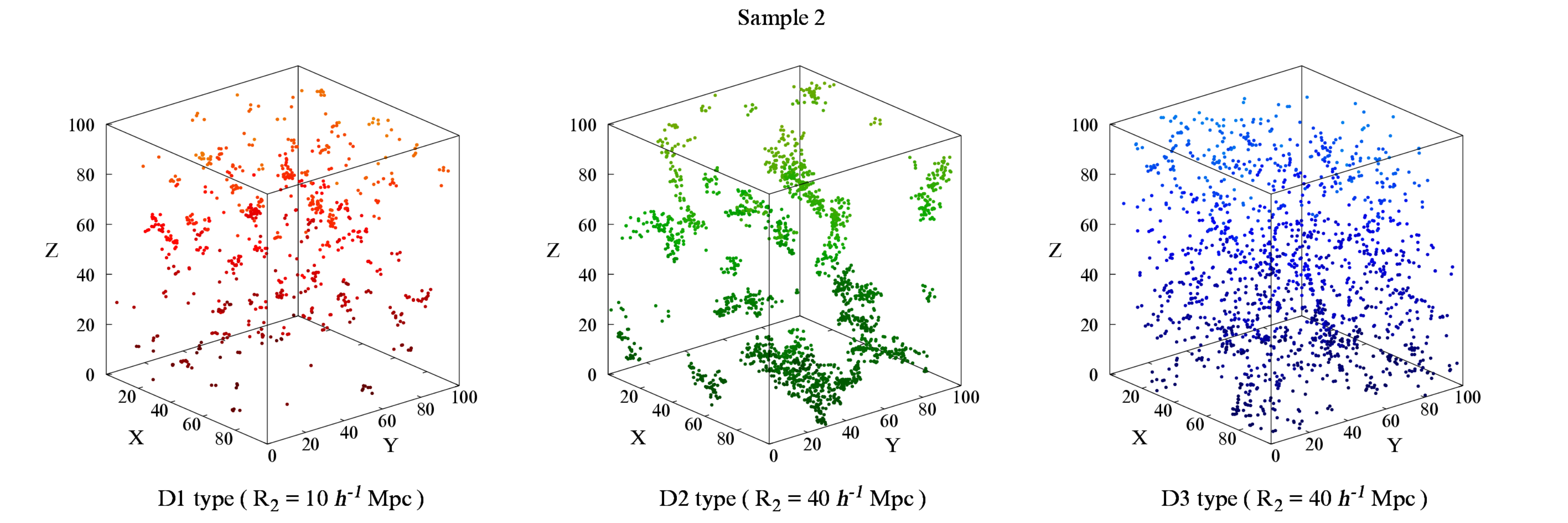}}}
\caption{The left, middle and right panels of this figure respectively
  show the distributions of $D1$ type, $D2$ type and $D3$ type
  galaxies within a cube of side $100 \hmpc$ from Sample 2. The $R_2$
  value used to calculate the local dimensions of galaxies in each
  case are mentioned at the bottom of the cubes. The gradients in the
  colours are applied according to the distances of the galaxies from
  the bottom of the cube.}
\label{fig:visual}
\end{figure*} 

\begin{figure*}
\resizebox{18 cm}{!}{\rotatebox{0}{\includegraphics{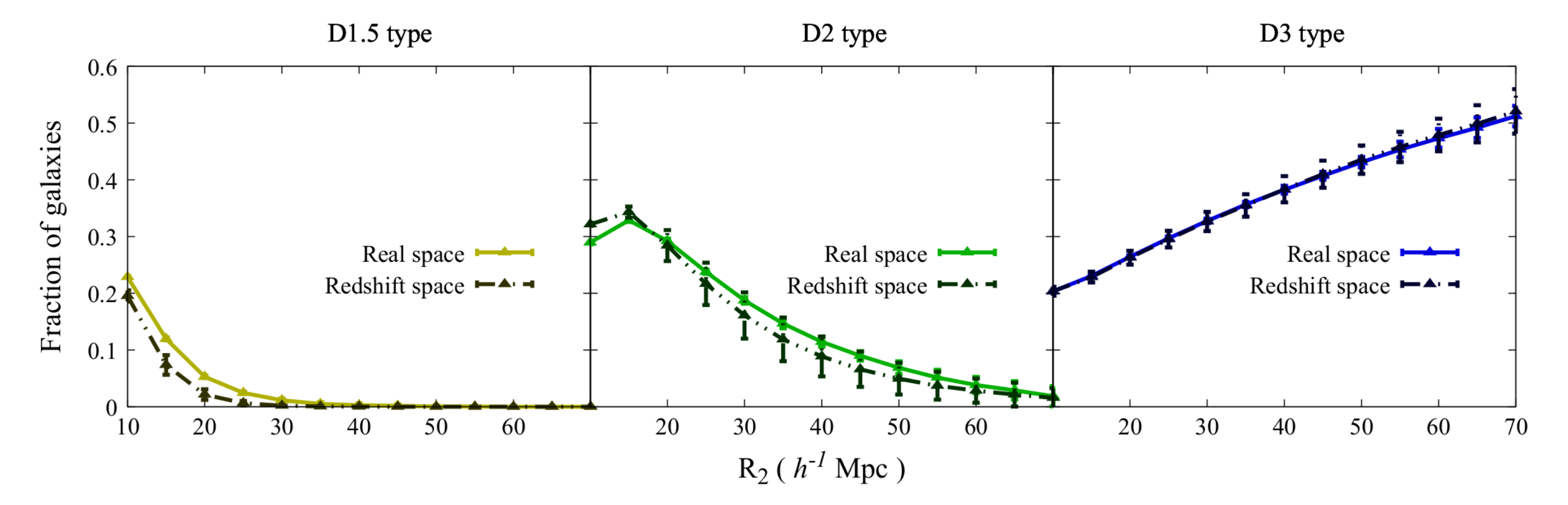}}}
\caption{The left, middle and right panel of this figure respectively
  show the fraction of $D1.5$, $D2$ and $D3$ type
  galaxies as a function of length scale in real and redshift
  space. The $1-\sigma$ error bars in each case are obtained from 8
  mock samples. Only the errors bars for redshift space are shown for
  clarity.}
\label{fig:rsd}
\end{figure*}

\begin{figure*}
\resizebox{18 cm}{!}{\rotatebox{0}{\includegraphics{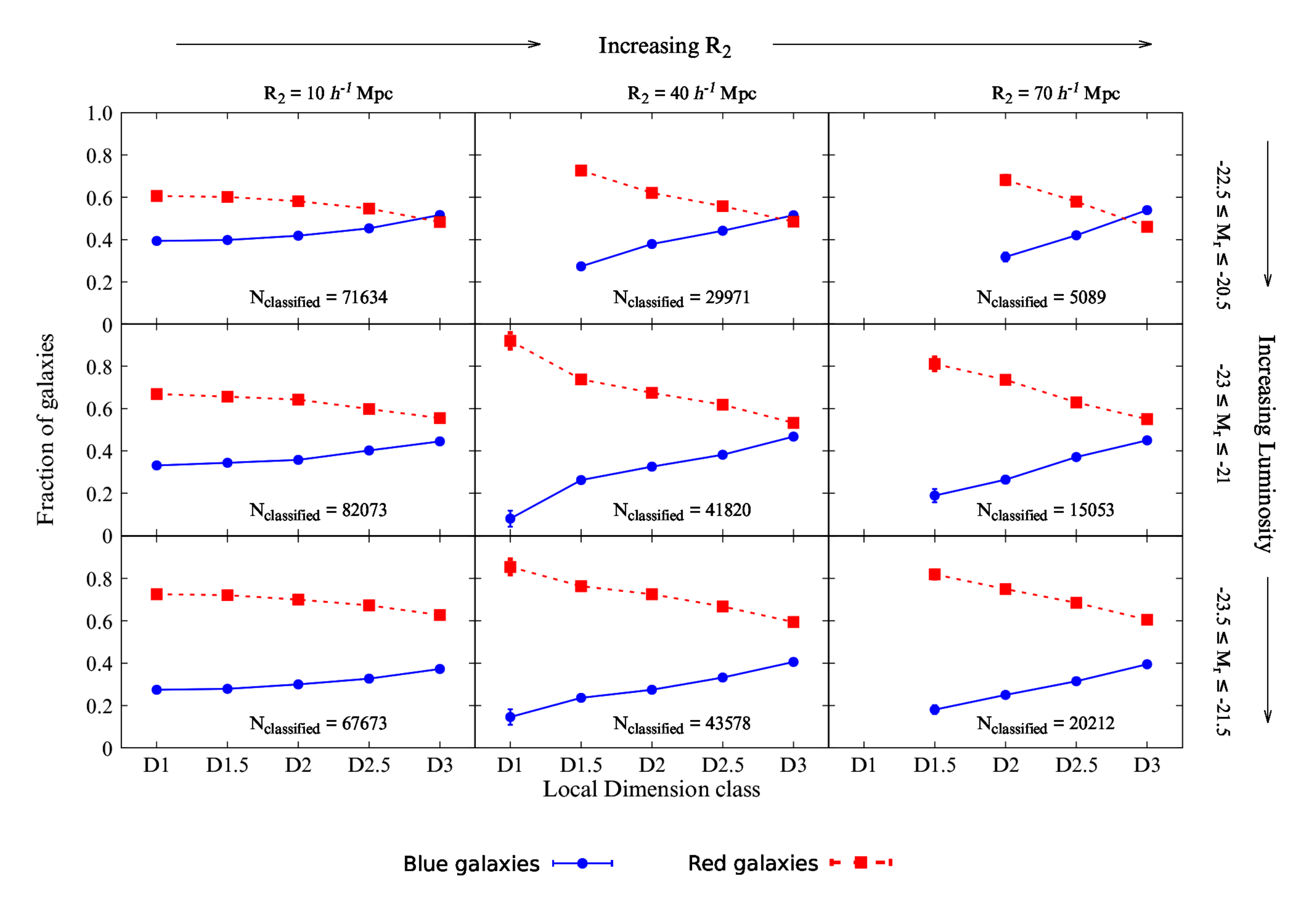}}}
\caption{ This figure shows the fraction of red and blue galaxies in
  different environments of the cosmic web for different length scales
  for Sample 1, Sample 2 and Sample 3 (\autoref{tab:samples}). The
  $1-\sigma$ error bars at each data point are calculated using $10$
  Jackknife samples drawn from each of the SDSS samples. The total
  number of classified galaxies at three different length scales for
  each of the samples are also mentioned in the respective panels. }
\label{fig:lpcd}
\end{figure*}

\begin{figure*}
\resizebox{18 cm}{!}{\rotatebox{0}{\includegraphics{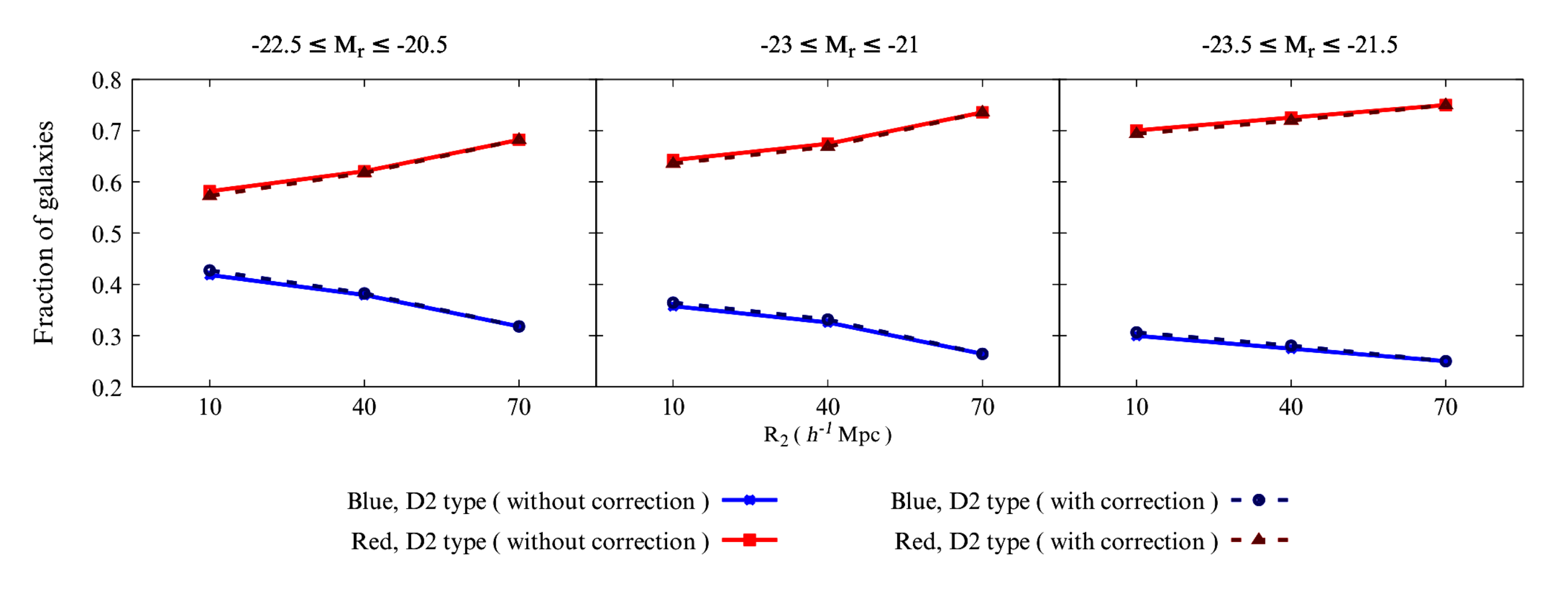}}}
\caption{This figure compares the corrected and uncorrected fraction
  of red and blue galaxies residing in sheets (D2 type) of different
  size in three volume limited samples (\autoref{tab:samples}). A
  galaxy can reside in a geometric environment with same local
  dimension on multiple length scales. In such cases, the galaxy is
  considered only on the largest among these length scales. The
  fraction of red and blue galaxies at different length scales are
  corrected by taking account the above mentioned possibility.}
\label{fig:D2diff}
\end{figure*}

\begin{figure*}
\resizebox{18 cm}{!}{\rotatebox{0}{\includegraphics{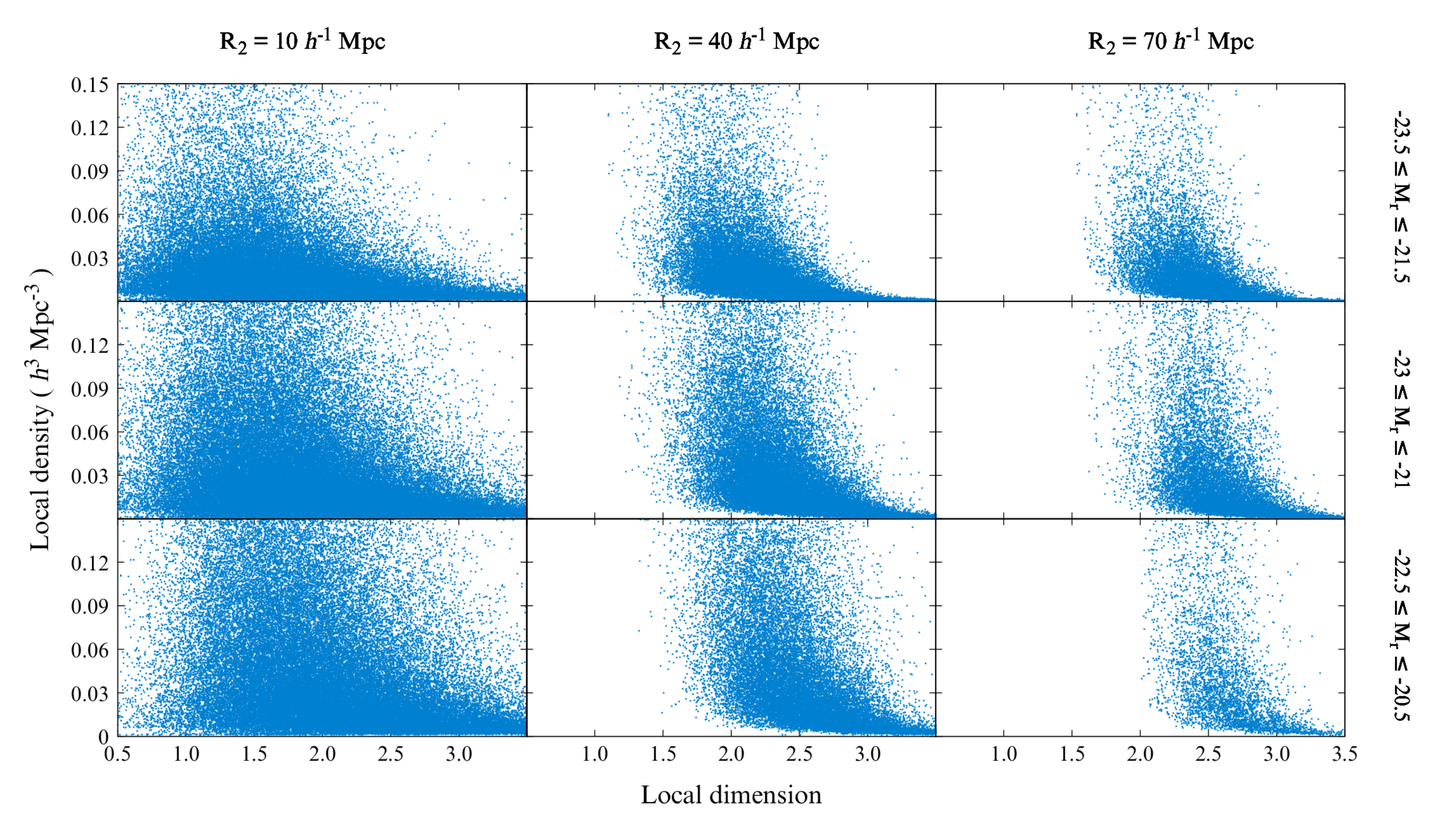}}}
\caption{Top, middle and bottom panels of this figure show local
  dimension vs local density in the three volume limited samples with
  different luminosity. The left, middle and right panels show results
  for different $R_2$ values.}
\label{fig:denld}
\end{figure*} 

\begin{figure*}
\resizebox{18 cm}{!}{\rotatebox{0}{\includegraphics{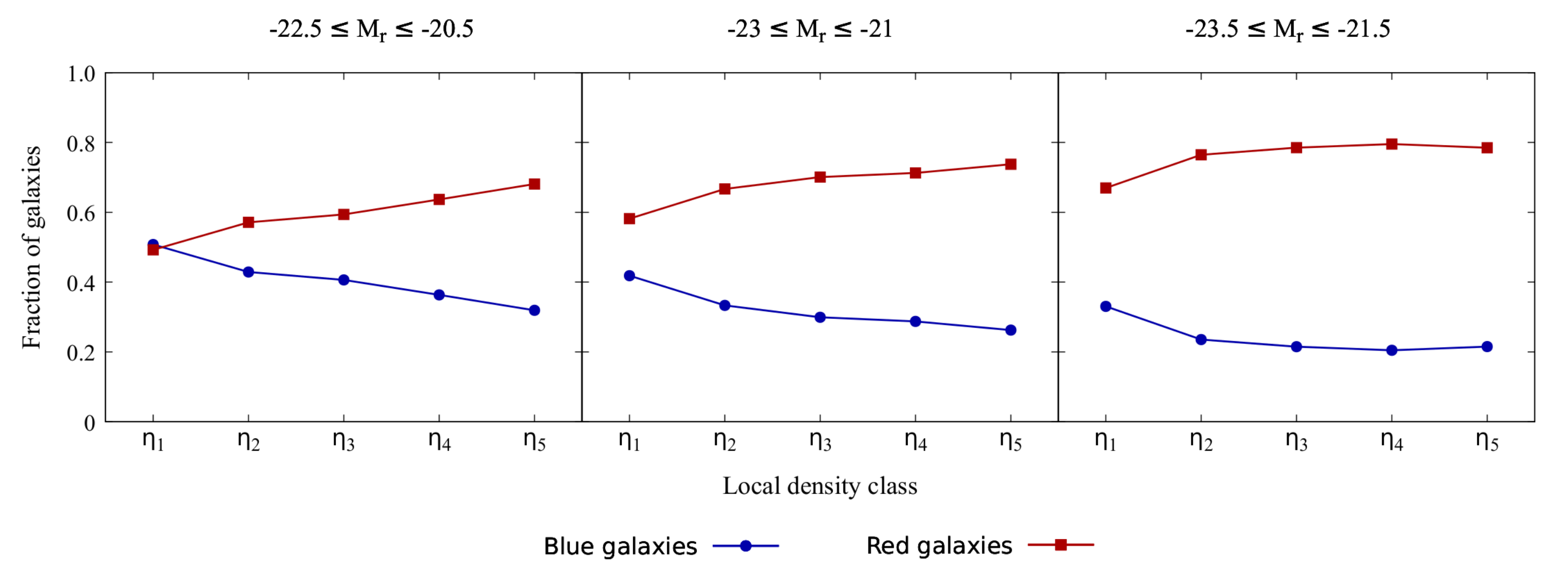}}}
\caption{The left, middle and right panels of this figure show the
  fraction of red and blue galaxies as a function of local density in
  the three volume limited samples.}
\label{fig:denfrac}
\end{figure*} 

\begin{figure*}
\resizebox{18 cm}{!}{\rotatebox{0}{\includegraphics{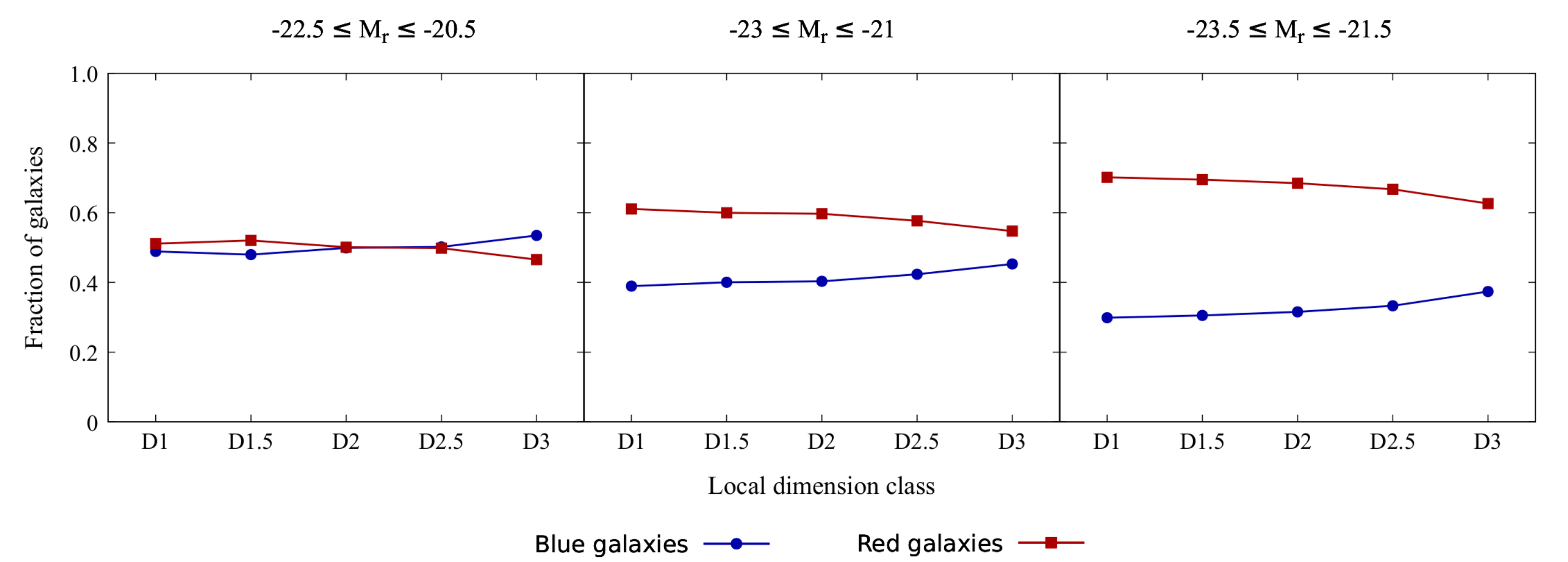}}}
\caption{This figure shows the fraction of red and blue galaxies as a
  function of local dimension at a fixed local density. The local
  dimensions shown in this figure are computed using $R_2=10
  \hmpc$. The left, middle and right panels of this figure corresponds
  to galaxies located at a fixed local density $\eta_1$ in the three
  volume limited samples.}
\label{fig:denmorph10}
\end{figure*} 

\begin{table*}{}
\caption{This table shows the number of classified galaxies in
  different geometric environments in the three volume limited samples
  (\autoref{tab:samples}) for three different values of $R_2$.}
\label{tab:classnum}
\begin{tabular}{cclrlrlr}
\hline
\hline
Sample  &	Total	 & \multicolumn{6}{c}{Number of galaxies classified at $R_2 \hmpc$}\\
Name		&	galaxies & \multicolumn{2}{c}{$R_2=10$}  & \multicolumn{2}{c}{$R_2=40$} & \multicolumn{2}{c}{$R_2=70$}\\
\hline
\multirow{6}{*}{Sample 1} & \multirow{6}{*}{90270} &  &	&	&	&	&		\\
 	& 	& Total: 		&71634 	& Total: 		&29971	& Total:		  	&5089	\\
 	&&&&&&& 	\\
 	&	& D1:  			&6362	& D1:  			&0		& D1:  			&0  	 	\\
 	&	& D1.5:  			&19229	& D1.5:  			&951		& D1.5:  			&0   	\\
 	&	& D2:  			&21060	& D2:  			&8464	& D2:  			&497		\\
 	&	& D2.5:  			&13213	& D2.5:  			&13037	& D2.5:  			&3100	\\
 	&	& D3:  			&11770	& D3:  			&7519	& D3:  			&1492	\\
 	&&&&&&& 	\\

\hline
\multirow{6}{*}{Sample 2} & \multirow{6}{*}{104137} &  &	&	&	&	&		\\
	&	& Total:			&82073	& Total:  		&41820 	& Total:  		&15053	\\
	&&&&&&& 	\\
 	&	& D1:  			&11167	& D1:  			&25		& D1:  			&0  	 	\\
 	&	& D1.5:  			&23630	& D1.5:  			&1483	& D1.5:  			&90   	\\
 	&	& D2:  			&22479	& D2:  			&12972	& D2:  			&1820	\\
 	&	& D2.5:  			&13207	& D2.5:  			&17749	& D2.5:  			&8370	\\
 	&	& D3:  			&11590	& D3:  			&9591	& D3:  			&4773	\\
 	&&&&&&& 	\\

\hline
\multirow{6}{*}{Sample 3} & \multirow{6}{*}{92848} &  &	&	&	&	&		\\
 	&	& Total:			&67673	& Total:			&43578	& Total:			&20212	\\
 	&&&&&&& 	\\
 	&	& D1:  			&12109	& D1:  			&55		& D1:  			&0  	 	\\
 	&	& D1.5:  			&20717	& D1.5:  			&2444	& D1.5:  			&144   	\\
 	&	& D2:  			&17177	& D2:  			&15700	& D2:  			&3999	\\
 	&	& D2.5:  			&9496	& D2.5:  			&17519	& D2.5:  			&11153	\\
 	&	& D3:  			&8174	& D3:  			&7860	& D3:  			&4916	\\
 	&&&&&&& 	\\
 	
\hline
\hline
\end{tabular}
\end{table*}

\begin{table*}{}
\caption{ This table shows the number of galaxies which are available
  at sheetlike environment on multiple length scales for the three
  volume limited samples.}
\label{tab:D2num}
\begin{tabular}{cccc}
\hline
\hline
Sample  &\multicolumn{3}{c}{Number of D2 type galaxies which are common at $R_2=$}\\
Name		&$10$ and $40\hmpc$ &$10$, $40$ and $70\hmpc$ &$40$ and $70\hmpc$ \\
\hline
Sample 1  & 2812 &  181 &  411   \\
Sample 2  & 3352 &  568 & 1219   \\
Sample 3  & 3126 & 1084 & 2890 	\\
\hline
\end{tabular}
\end{table*}

\begin{figure*}
\resizebox{18 cm}{!}{\rotatebox{0}{\includegraphics{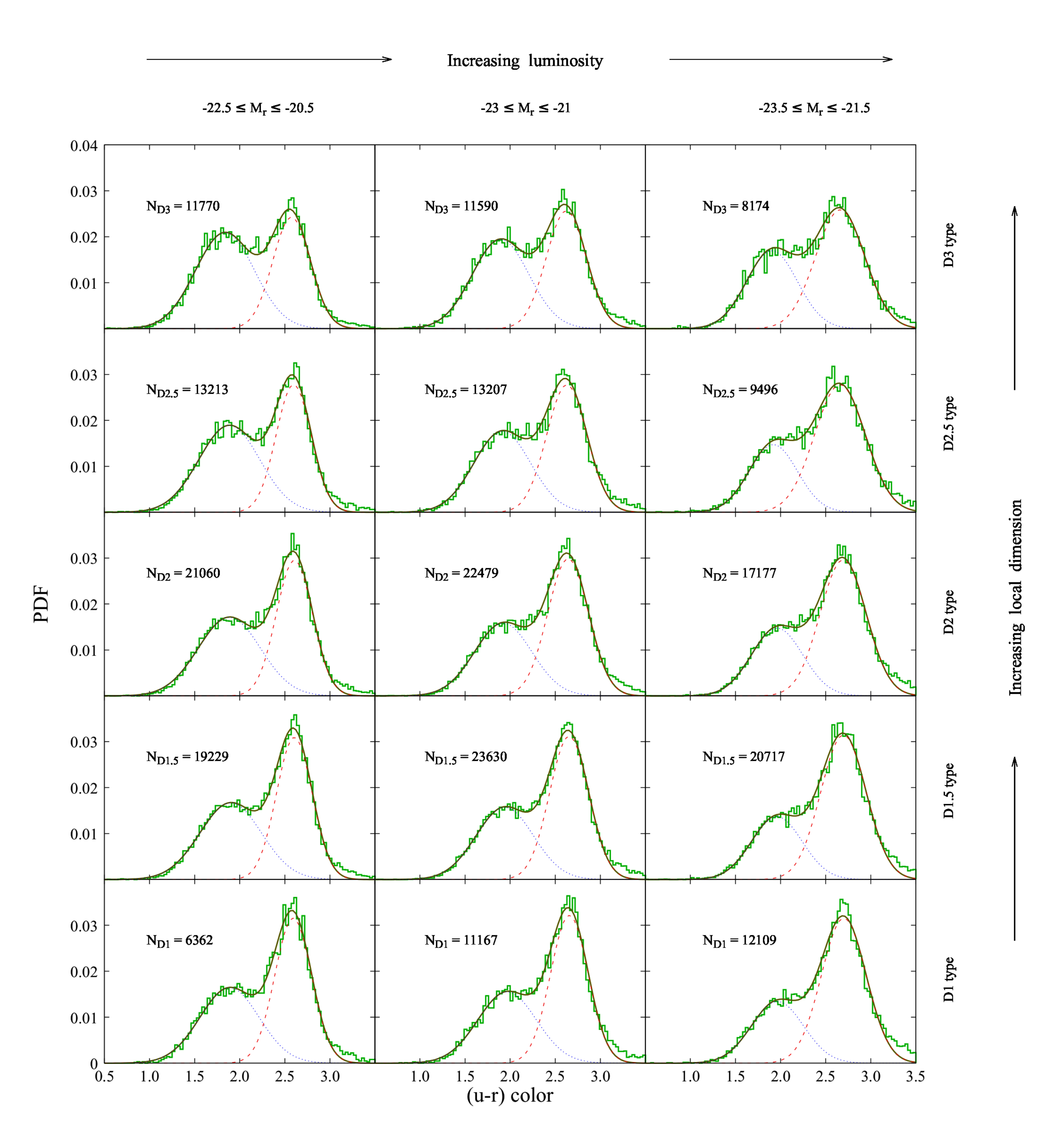}}}
\caption{The different panels of this figure shows the distribution of
  observed $u-r$ colour at different types of geometric environment on
  $10 \hmpc$ in three volume limited samples. The total number of
  galaxies identified at different environments on $10 \hmpc$ for each
  samples are mentioned in the respective panels. The smooth solid
  line in each panel represent the best fit double Gaussian describing
  the observed $u-r$ colour distribution whereas the dotted and dashed
  lines show the two individual components of the double Gaussian
  distribution.}
\label{fig:fitdg}
\end{figure*}

\begin{figure*}
\resizebox{18 cm}{!}{\rotatebox{0}{\includegraphics{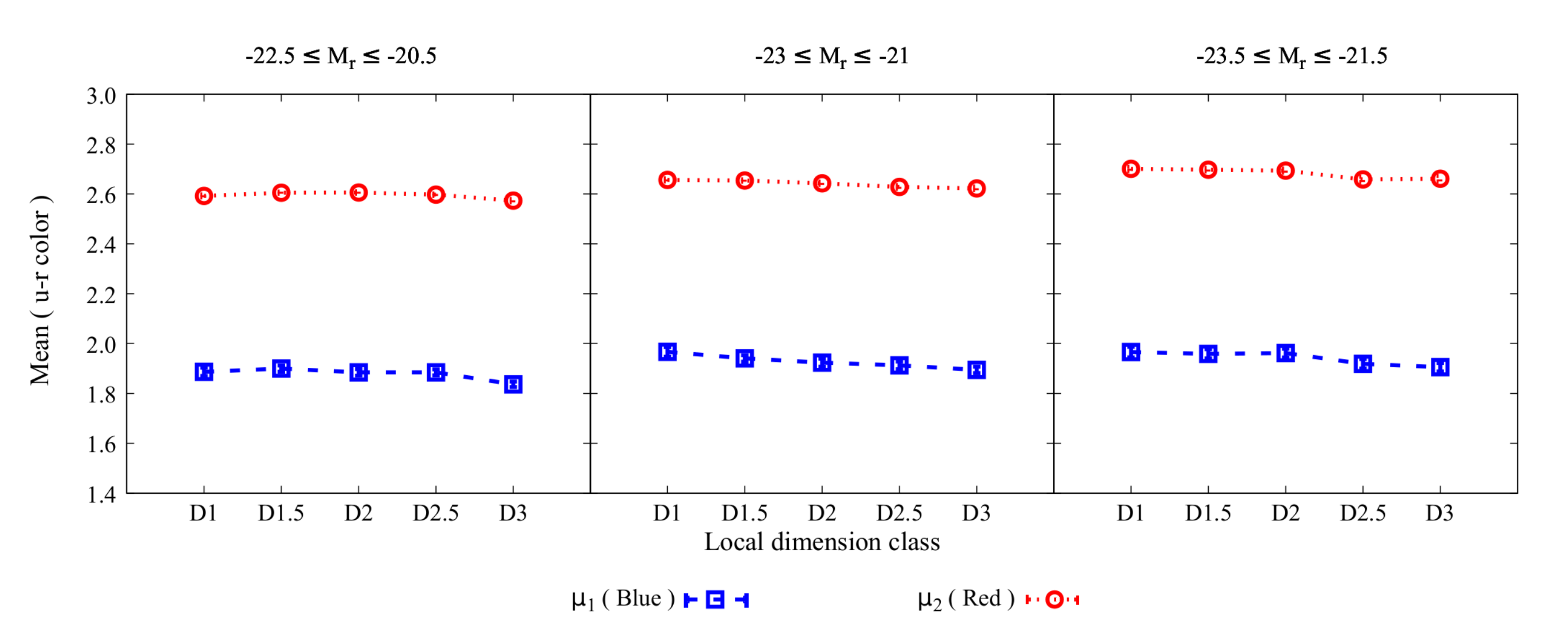}}}
\caption{This figure shows the best fit values of the means $\mu_1$
  and $\mu_2$ in \autoref{eq:fitfunc} as a function of local dimension
  in three volume limited samples. We fit the double Gaussian
  (\autoref{eq:fitfunc}) to the observed $u-r$ colour
  distribution. The errors shown at the data points are the standard
  errors for the fitted values of mean. The standard errors are very
  small here which are hardly visible in this plot.}
\label{fig:fitmean}
\end{figure*}

\begin{figure*}
\resizebox{18 cm}{!}{\rotatebox{0}{\includegraphics{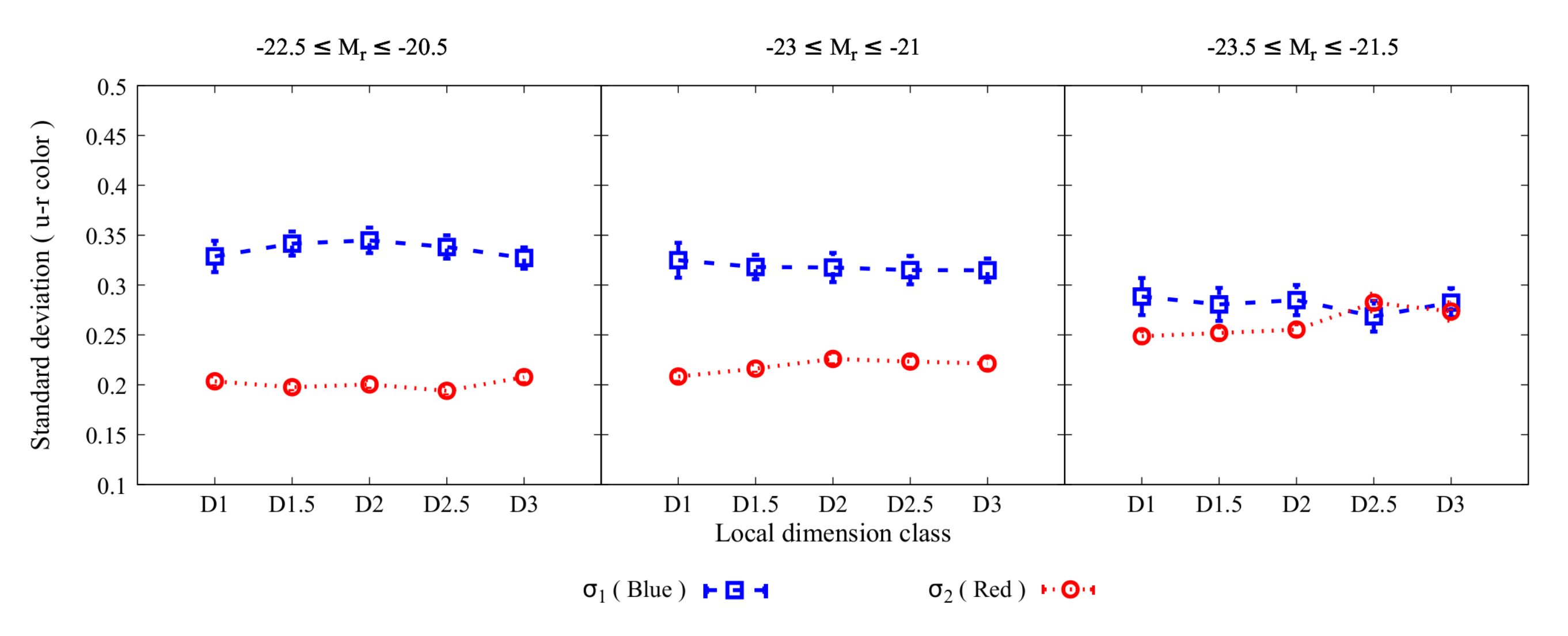}}}
\caption{This figure shows the best fit values of the standard
  deviations $\sigma_1$ and $\sigma_2$ in \autoref{eq:fitfunc} as a
  function of local dimension in three volume limited samples. The
  errors shown at the data points are the standard errors for the
  fitted values of dispersion.}
\label{fig:fitstd}
\end{figure*}

\section{DATA}
\subsection{SDSS data}
 The Sloan Digital sky survey (SDSS) is a multi band imaging and
 spectroscopic redshift survey which covers more than one third of the
 celestial sphere. We use the Main galaxy sample for the present
 analysis. The spectroscopic target selection for the Main galaxy
 sample is described in \citet{strauss02}. The photometric camera used
 in SDSS is described in \citet{gunn}.  \citet{fuku} describe the
 photometric system used in the survey. The collection of data for the
 Main Galaxy sample was completed by SDSS DR7 \citep{abaz}. We use
 data from the $16^{th}$ data release \citep{ahumada} of the Sloan
 Digital sky survey (SDSS). SDSS DR16 is the fourth and final data
 release of SDSS IV which incorporates data from all the prior data
 releases. We download the data from the SDSS
 SkyServer \footnote{https://skyserver.sdss.org/casjobs/} using
 Structured Query Language (SQL). For the present analysis, we select
 a contiguous region of the sky which spans $0^{\circ} \leq \delta
 \leq 60^{\circ}$ and $ 135^{\circ} \leq \alpha \leq 225^{\circ}$
 where $\alpha$ and $\delta$ are the right ascension and declination
 respectively. We select all the galaxies with r-band Petrosian
 magnitude $ 13.5 \le r_{p} < 17.77$ and redshift $z < 0.3$. These
 cuts provides us a total $376495$ galaxies. We then construct three
 volume limited samples (\autoref{fig:zM}) by applying cuts to the
 K-corrected and extinction corrected $r$-band absolute magnitude. The
 properties of these volume limited samples are described in detail in
 \autoref{tab:samples}.

We apply the method described in the previous section to identify the
galaxies in different environments in each of these volume limited
samples. The exact number of galaxies classified in different classes
at three different values of $R_2$ are tabulated in
\autoref{tab:classnum} for the three volume limited samples.

\subsection{MultiDark simulation data}
The three dimensional distribution of SDSS galaxies are obtained from
their spectroscopic redshifts. The redshifts are perturbed by peculiar
velocities of galaxies which distort their clustering pattern in
redshift space \citep{kaiser}. We would like to quantify the effects
of redshift space distortions on local dimension using mock samples
from N-body simulations. We prepare a set of mock samples for the
Sample 2 (\autoref{tab:samples}) using data from the cosmosim
project\footnote{https://www.cosmosim.org/}.  The MultiDark Planck 2
(MDPL2) simulation is part of a number of simulations \citep{klypin16}
based on cosmological parameters from Planck. The simulation is
executed with $3840^{3}$ dark matter particles in a cube of size $1000
\hmpc$, starting from redshift $120$. The mass resolution of the
simulation is $1.5\times 10^{9} h^{-1} M_{\odot}$.  The values of the
cosmological parameters used in the simulation are $H_0 = 67.77$,
$\Omega_m = 0.307$, $\Omega_{\Lambda}= 0.693$, $\Omega_{b}=0.048$,
$n=0.96$ and $\sigma_8 =0.82$. We have used the ROCKSTAR catalogue
\citep{behroozi13} of MDPL2. We use a SQL query to retreive the
position and velocities of the dark matter particles in snapshot 125
($z=0$). We place the observer at each corners of the cube and
randomly select $104137$ particles from each of these non-overlapping
regions to prepare 8 mock samples which have exactly identical
geometry as Sample 2. We prepare the mock samples both in real space
and redshift space. The mock samples in redshift space are constructed
by computing redshifts of galaxies from their distances in real-space
and peculiar velocities.  We analyze these mock samples in the same
way as the actual data .

\section{Results}
\subsection{Variations of environments with length scale}

The different environments of the cosmic web have different
characteristic size and their abundance would depend on the length
scales probed. We first quantify the number of classifiable galaxies
in each type of environment. The top three panels of
\autoref{fig:frac} show the number of classified galaxies in different
types of environment as a function of length scale in Sample 1, Sample
2 and Sample 3. The respective fraction of galaxies available at each
type of environment on each length scale for the three volume limited
samples are shown in the bottom three panels of the same figure. We
choose $R_1=5 \hmpc$ and then increase $R_2$ in uniform steps of $5
\hmpc$ starting from $R_2=10 \hmpc$. The number of classifiable
galaxies in each sample is expected to decrease with increasing $R_2$
due to their finite size.

In the bottom three panels of \autoref{fig:frac}, we find that the
galaxies live in diverse environments on small scales. In each of the
three volume limited samples, nearly $\sim 60-70\%$ galaxies reside in
filaments and sheets when the environment is characterized within a
length scale of $10-30 \hmpc$. The $D1$ type environments are mostly
associated with the straight filaments which extend only up to $\sim
30 \hmpc$ in these samples. The galaxies which are part of the curved
or warped filaments would be mostly represented by the $D1.5$ type
galaxies. We find that the $D1.5$ type environment extends up to $\sim
50 \hmpc$. The fraction of $D1$ and $D1.5$ type galaxies steadily
decrease with increasing length scales. The $D2$ type galaxies reside
in sheets which are the most abundant environment within a scale of
$\sim 30 \hmpc$. The abundance of sheet peaks around $20-30 \hmpc$ in
all three volume limited samples. The fraction of galaxies in
sheetlike environment decreases with increasing length scales and only
$\sim 10-20\%$ galaxies are part of sheetlike structures at $70
\hmpc$. No galaxies are found to be a part of filamentary environment
at this length scale. Comparison of the results from the three volume
limited samples suggests that the filamentary and sheetlike
environments can be traced to a slightly larger length scales in the
brighter samples. This results from the larger number of classifiable
galaxies available at larger length scales due to the bigger volumes
of the brighter samples. At $70 \hmpc$, $80-90\%$ of the galaxies are
part of $D2.5$ and $D3$ type environments which indicates that the
diversity of environment ceases to exist on larger length scales. The
fact that the local dimension of galaxies shift towards $3$ hints
towards the emergence of a homogeneous network of galaxies on
sufficiently large length scales. This is consistent with the findings
of various studies on large-scale homogeneity which suggests that the
Universe is statistically homogeneous on scales beyond $\sim
100-150\hmpc$ \citep{yadav, hogg, sarkarp, scrim, nadathur, pandey15,
  pandey16, avila}. In \autoref{fig:visual}, we separately show the
distributions of $D1$, $D2$ and $D3$ type galaxies within a cube of
side $100 \hmpc$. The left panel of \autoref{fig:visual} shows that
$D1$ type galaxies mostly reside in short and straight filaments.
The middle panel shows that $D2$ type galaxies occupy sheetlike
environments which are visually most prominent. The distribution of
$D3$ type galaxies in the right panel is nearly homogeneous in nature.

\subsection{Effects of redshift space distortions on local dimension}
We analyze mock SDSS samples from the MultiDark simulations in real
space and redshift space to quantify the effects of redshift space
distortions on local dimension. We compute the fraction of $D1.5$,
$D2$ and $D3$ type galaxies in the mock SDSS samples as a function of
length scale $R_2$ in real space and redshift space. Only a small
fraction of galaxies are found to reside in $D1$ type environments in
these mock samples. So we do not show the results for $D1$ type
galaxies here. We compare the results in real space and redshift space
in each case. The results are shown in \autoref{fig:rsd}. The left,
middle and right panel of \autoref{fig:rsd} respectively show the
fraction of $D1.5$, $D2$ and $D3$ type galaxies as a function of
length scale in real and redshift space. The results in redshift space
are qualitatively similar to what we observe for SDSS galaxies in
\autoref{fig:frac}. The left panel of this figure suggests that there
is a marginal decrease in the fraction of $D1.5$ type galaxies in
redshift space. The same trend is also observed for $D1$ type
galaxies. This result is somewhat counter intuitive as one would
expect an enhancement of short filaments in redshift space due to the
`Finger of God'(FOG) effect. The `FOG' stretches the dense virialized
groups along the line of sight which are expected to resemble short
filaments. However, we do not observe any such trend with $D1$ or
$D1.5$ type galaxies in our mock SDSS sample. This may happen if the
analyzed sample do not contain sufficient virialized dense groups in
it. Some of the short filaments with orientations other than parallel
and perpendicular to the line of sight in real space, may not be
identified as filaments in redshift space, which causes a marginal
decrease in their abundance. In the middle panel, there is a marginal
increase in the fraction of $D2$ type galaxies in redshift space on
small scales. The trend reverses on large scales showing a small
decrease in the fraction of $D2$ type galaxies in redshift
space. However, these differences are well within the $1-\sigma$ error
bars or very close to it. The right panel of \autoref{fig:rsd} shows
that the fraction of $D3$ type galaxies are least affected by redshift
space distortions. We also repeat the same analysis with mock samples
for Sample 3 and find analogous results. The test with the mock
samples suggests that the redshift space distortions do not impart a
statistically significant influence on the local dimension of galaxies
and there are no preferred scale in the analysis.

\subsection{Variations of red and blue fractions with geometric environment and length scale}
We show the fraction of red and blue galaxies in different
environments of the cosmic web for Sample 1, Sample 2 and Sample 3
respectively in the top, middle and bottom panels of
\autoref{fig:lpcd}. The left, middle and right panels at the top,
middle and bottom row of \autoref{fig:lpcd} corresponds to three
different length scales $10 \hmpc$, $40 \hmpc$ and $70 \hmpc$
respectively. Thus each row in \autoref{fig:lpcd} corresponds to a
fixed luminosity and each column corresponds to a a fixed length
scale. The \autoref{fig:lpcd} shows that environments with smaller
local dimension tend to host a higher fraction of red galaxy. The red
fraction at each local dimension increases when such environments are
traced on larger length scales. There is also a clear increase in the
red fraction at each environment with increasing luminosity. The
luminosity dependence of red fraction is a well known phenomena. So we
do not discuss the related results in greater detail. We only focus
our attention on the dependence of red fraction on geometric
environment and the length scales associated with such environment. We
find that fraction of red galaxies dominates over the blue fraction at
nearly each environment and each length scale. This is not necessarily
true for all galaxy samples. We find that a reverse trend exists in
fainter galaxy samples which are not included in this
analysis. However, the decrease of red fraction with increasing local
dimension is a common trend in all galaxy samples irrespective of
their luminosity.

The \autoref{fig:lpcd} shows that the fraction of red and blue
galaxies also depends on the length scales associated with the
environment. At a fixed luminosity and a fixed environment, the
fraction of red galaxies increases with increasing length scales. We
show the fraction of red and blue galaxies residing in sheets as a
function of their size in the first 3 volume limited samples in
\autoref{fig:D2diff}. The figure clearly shows that there is an
increase in the fraction of red galaxies and a decrease in the
fraction of blue galaxies with the increasing size of these
structures. It is worthwhile to mention here that a subset of galaxies
can be identified in the same type of environment on multiple length
scales. These galaxies are part of such environment which extends at
least up to the largest among these length scales. This must be taken
into account while estimating the fraction of red or blue populations
in different environments. We quantify the number of such galaxies at
each environment for all the samples. For example, the number of such
galaxies in $D2$ type environment for different length scales are
tabulated in \autoref{tab:D2num}. This table shows that in Sample 1,
there are $181$ galaxies which reside in sheets at all three length
scales i.e. $10 \hmpc$, $40 \hmpc$ and $70 \hmpc$. These $181$
galaxies are part of sheets extending to at least $70 \hmpc$. Further,
$2812$ galaxies are part of sheets both at $10 \hmpc$ and $40 \hmpc$
but not at $70 \hmpc$. These galaxies are part of sheets which are
extended up to $40 \hmpc$. We subtract these two numbers from the total
number of galaxies identified in sheets on $10 \hmpc$. Similarly,
$411$ galaxies are identified in $D2$ type environment both at $40
\hmpc$ and $70 \hmpc$ in Sample 1. The sheets associated with these
galaxies extend at least up to $70 \hmpc$. We subtract this number from
the total number of galaxies found in sheets on a length scale of $40
\hmpc$. The red and blue fractions in $D2$ type environment are also
corrected in Sample 2 and Sample 3 in a similar manner. We show both
the corrected and the uncorrected fractions of red and blue galaxies
in sheetlike environment at different length scales in
\autoref{fig:D2diff}. We find that these corrections hardly make any
difference to these results.

\subsection{Comparing effects of density and geometry of environments on the red and blue fractions}
Galaxy colour is known to be sensitive to local density. Generally
sheets are denser than fields and filaments are denser than sheets. So
the dependence of red and blue fractions on the geometry of
large-scale environment shown in \autoref{fig:lpcd} may partly arise
due to dependence of galaxy colour on local density. We need to
decorrelate the effect of local density in order to test the role of
large scale structures on galaxy colours. We address this issue by
calculating local number density of galaxies using $k^{th}$ nearest
neighbour method \citep{casertano85}. We compute the local number
density using \autoref{eq:knn} with $k=5$. We plot the local density
against the local dimension of galaxies in three different volume
limited samples in \autoref{fig:denld}. The top left, middle left and
bottom left panels of \autoref{fig:denld} show the relations between
local dimension and local density for the three volume limited samples
when local dimensions are computed using $R_2=10 \hmpc$. The results
in these panels show that the environments with larger local dimension
indeed tend to have a lower local density. However these relationships
show very large scatters. The environments with local dimension up to
$D=2.5$ can have a wide range of local densities and it is difficult
to assign a specific density range to the environments with different
local dimension. The three panels in the middle column and three
panels in the right column of \autoref{fig:denld} show the relations
between local density and local dimension in these samples for $R_2=40
\hmpc$ and $R_2=70 \hmpc$. They show a similar trend as seen in the
three panels in the left column of the same figure. It may be noted
that galaxies with smaller local dimension are progressively absent
when the geometry of environments are characterized on larger length
scales. This points out to the emergence of a homogeneous network of
galaxies on larger length scales as mentioned earlier.

We first study the effect of local density on the red and blue
fractions. We divide the galaxies into five different density classes
based on their local density values. The different density classes are
defined in \autoref{tab:nnden}. The left, middle and right panels of
\autoref{fig:denfrac} show the red and blue fractions of galaxies for
different density classes in three volume limited samples. The left
panel shows a clear increase in red fraction with increasing
density. A higher fraction of red galaxies are observed for the
brighter samples in the middle and right panels of
\autoref{fig:denfrac}. However the density dependence are less
pronounced in the brighter samples which indicates that the brighter
galaxies are intrinsically redder irrespective of the density of their
environments. This results may be related to the fact that brighter
galaxies have higher stellar mass, which are mostly red in all
environments \citep{bamford}. Galaxies with lower luminosity have low
stellar mass which are known to be blue in low density environment and
red in high density environment. Our results are in good agreement
with \citet{bamford}.

We then compare the red and blue fractions across the environments
with different local dimension but at a fixed local
density. \autoref{fig:denld} shows that at $R_2=10 \hmpc$, the
lowermost density class $\eta_1$ has an uniform coverage of local
dimension for all three volume limited samples. We consider all the
galaxies in this density class and plot the red and blue fractions at
different geometric environments for the three volume limited samples
in \autoref{fig:denmorph10}.  Interestingly, the red fractions still
show a mild decrease with increasing local dimension in each of the
three panels of this figure. This suggests that geometry of
environment also plays a role in deciding galaxy colour besides the
local density. Unfortunately, we can not carry out this test for all
density classes and all $R_2$ values due to non-uniform coverage
(\autoref{fig:denld}).

 This analysis indicates that both local density and geometry of
 environments play a role in deciding the colours of galaxies. The
 geometry and density dependence of red and blue fractions shown in
 \autoref{fig:lpcd} and \autoref{fig:denfrac} are most likely an
 outcome of the combined effects of density and geometry which are
 difficult to disentangle.
 
\subsection{Bimodality of the colour distribution in different environments and luminosity}

Galaxy colour is known to follow a bimodal distribution. We fit the
distribution of the $u-r$ colour of galaxies using a double Gaussian
with a PDF,
\begin{eqnarray}
f(x)&=& A_1 \exp{\left[\frac{(x- \mu_1)^2}{2 \sigma_1^2}\right]}+A_2 \exp{\left[\frac{(x- \mu_2)^2}{2 \sigma_2^2}\right]} 
\label{eq:fitfunc}
\end{eqnarray}
where $A_1=\frac{\alpha_1}{\sqrt{2\pi} \sigma_1}$ and
$A_2=\frac{\alpha_2}{\sqrt{2\pi} \sigma_2}$ are the amplitudes of the
two Gaussians. $\alpha_1$, $\alpha_2$ are the associated weights,
$\mu_1$, $\mu_2$ are the means and $\sigma_1$, $\sigma_2$ are the
standard deviations of the two Gaussian components respectively.  The
usefulness of a double Gaussian fit in modelling galaxy colour
distribution has been explored in a number of previous works
\citep{balogh2004, baldry2006}. This approach has an inherent
limitation due to the significant overlap of the two Gaussian
distributions. A more rigorous approach to overcome this limitation is
presented in \citet{taylor15}.

We plot the distributions of $u-r$ colour in different types of
geometric environments on a length scale of $10 \hmpc$ for the three
volume limited samples in different panels of \autoref{fig:fitdg}. We
fit each of these distributions using a two Gaussian distribution
(\autoref{eq:fitfunc}). In each panel, the left and right peaks of the
bimodal distribution corresponds to the blue and red galaxies
respectively. We can clearly see that at each luminosity, the
distribution of red galaxies are more sharply peaked than the blue
galaxies in the geometric environments with smaller local dimension.

The mean and dispersion of the distributions at each geometric
environment for the three volume limited samples are shown in
\autoref{fig:fitmean} and \autoref{fig:fitstd} respectively. The
\autoref{fig:fitmean} shows that the mean colour of both the red and
blue populations mildly decrease with increasing local dimension. At
each geometric environment, the mean colour increases with increasing
luminosity. We do not find a clear environmental trend for the
dispersions in \autoref{fig:fitstd}. The dispersion for the red
population is smaller than that for the blue population at each
geometric environment. This is consistent with the visual impression
from \autoref{fig:fitdg} mentioned earlier. The dispersions for the
red and blue populations at each geometric environment respectively
increases and decreases with increasing luminosity.

The results of this analysis show that red galaxies prefer to reside
in geometric environments with smaller local dimension. More luminous
galaxies are found to be redder as expected.  These findings are
consistent with \autoref{fig:lpcd} which show a similar dependence of
the red and blue fraction on the local dimension and
luminosity. Despite these variations across different environments and
luminosities, the valley which separates the blue and red
distributions appear around the same colour cut $(u-r)=2.22$ used to
separate the two distributions. Finally, \autoref{fig:fitdg} shows
that the bimodal nature of the colour distribution is present across
all environments and luminosities.

 We could not repeat this analysis for $R_2=40 \hmpc$ and $R_2=70
 \hmpc$ due to a relatively smaller number of galaxies available at
 these environments on those length scales.


\section{CONCLUSIONS}
We study the dependence of galaxy colour on different environments of
the cosmic web using the data from SDSS DR16. We analyze a number of
volume limited samples with different luminosity. We identify the
galaxies in different environments by quantifying their local
dimension on different length scales and estimate the fraction of red
and blue galaxies in each environment for each of these samples. The
analysis shows that for a fixed length scale, the fraction of red
galaxies decreases with the increasing local dimension at each
luminosity. At a fixed length scale, the fraction of red galaxies in
each environment also increases with luminosity. The environments with
lower values of local dimension are expected to represent denser
regions of the cosmic web. Also the more luminous galaxies are known
to be hosted in higher density regions \citep{einasto03,
  berlind}. These findings are consistent with the earlier studies on
the environmental dependence of galaxy colour \citep{hogg2004,
  baldry2004, blan2, ball2008, bamford}. The observed fraction of red
galaxies thus depends on several interdependent parameters such as
local density, geometry of environment and luminosity of galaxies. It
is in general difficult to separate the role of these parameters in
influencing colours of galaxies.

We separately study the dependence of red fractions on local density
and find that the fraction of red galaxies increases with density. The
trend persists across all the luminosity bins. The brighter samples
host a higher fraction of red galaxies at the same density. At a fixed
density, we study the fraction of red galaxies as a function of local
dimension. This helps us to decorrelate the effect of density and test
the effect of geometric environments on galaxy colours. We find that
the fraction of red galaxies still decreases with increasing local
dimension which suggests that geometry of environments also plays a
role in determining the colours of galaxies. The effects of density
and geometry are coupled due to their mutual correlation which is
difficult to separate. A part of the density dependence of galaxy
colours may arise due to geometry of their environment and vice versa.

The present analysis do not identify separate structures but only
determine the geometric environment around a galaxy. We estimate the
local dimension on various length scale ranges and consider only the
best quality fits. So the estimated local dimension on a given length
scale characterize the geometry of the environment of a galaxy on that
length scale. We can not establish a direct relationship between the
$R_2$ values and the size of the environment but it can be used as a
proxy for their extent due to the involvement of the length scale in
the fitting method for local dimension \citep{sarkar19}. At a fixed
luminosity and a fixed local dimension, the fraction of red galaxies
increases with $R_2$. This indicates that the filaments and sheets
which extends to larger length scales, host a higher fraction of red
galaxies and a lower fraction of blue galaxies than their smaller
counterparts. The relative change in the red fraction with the length
scales is more pronounced in the fainter samples than the brighter
samples. This is related to the fact that the effects of luminosity
dominates in the brighter samples. The filaments and sheets extending
to larger length scales are expected to have a larger baryon reservoir
which must have started to form earlier \citep{zeldovich}. These
together would favour a higher accretion rate and a larger stellar
mass for the galaxies in these environments. A higher accretion rate
sustained for a longer period would exhaust the supply of gas in the
surrounding environment leading to a higher fraction of red
galaxies. It may be noted that a significant fraction of the red
galaxies are known to be spirals \citep{masters} which reside in
filamentary and sheetlike environments of the cosmic web.

The distribution of galaxy colour in each environment can be described
by a double Gaussian. We find that the mean colour of galaxies becomes
redder with decreasing local dimension of the host environment and
increasing luminosity of the sample. The dispersions of the red and
blue distributions do not exhibit a clear trend with geometric
environment. The results indicate that the environments with lower
local dimension favour the transformation from blue to red
galaxies. However the bimodal nature of the colour distribution
persists in all environments which suggests that such transformations
are allowed in all environments possibly via different physical
mechanisms.

Finally, we would like to emphasize that the colour of a galaxy depends
on the size of its host environment besides density and
luminosity. The present analysis shows that the larger structures
contain a higher fraction of red galaxies and a lower fraction of blue
galaxies. Superclusters are known to have a sheetlike morphology with
an intervening filamentary environment \citep{costa, einasto11,
  einasto17}. \citet{einasto14} reported a higher fraction of red
galaxies in filament-type superclusters. It would be also interesting
to study the abundance and distribution of the red and blue galaxies
in the individual superclusters of different size. Further, the recent
studies with the Galaxy Zoo reveal that a large number of red galaxies
are massive spirals \citep{bamford, masters, tojeiro}. \citet{masters}
find that the local density alone is not sufficient to explain the
colour of these galaxies. They reported that the massive galaxies are
red independent of their morphology.

The results of the present analysis suggests that if the massive red
spirals are hosted in larger filaments or sheets then their colour may
be explained by their embedding large-scale environment which favours
a higher accretion rate leading to a larger stellar mass and
consequently a faster depletion of the surrounding gas reservoir. In
future, we plan to carry out such an analysis with the Galaxy Zoo
which would help us to better understand the role of the embedding
large-scale structures on the galaxy formation and evolution.

 \section{ACKNOWLEDGEMENT}
The authors thank an anonymous reviewer for useful comments and
suggestions which helped us to improve the draft.  BP would like to
acknowledge financial support from the SERB, DST, Government of India
through the project CRG/2019/001110. BP would also like to acknowledge
IUCAA, Pune for providing support through associateship programme. SS
would like to thank UGC, Government of India for providing financial
support through a Rajiv Gandhi National Fellowship. SS would also like
to acknowledge Biswajit Das for useful discussions.

The authors would like to thank the SDSS team for making the data
public.  Funding for the Sloan Digital Sky Survey IV has been provided
by the Alfred P. Sloan Foundation, the U.S. Department of Energy
Office of Science, and the Participating Institutions. SDSS-IV
acknowledges support and resources from the Center for
High-Performance Computing at the University of Utah. The SDSS web
site is www.sdss.org.

SDSS-IV is managed by the Astrophysical Research Consortium for the
Participating Institutions of the SDSS Collaboration including the
Brazilian Participation Group, the Carnegie Institution for Science,
Carnegie Mellon University, the Chilean Participation Group, the
French Participation Group, Harvard-Smithsonian Center for
Astrophysics, Instituto de Astrof\'isica de Canarias, The Johns
Hopkins University, Kavli Institute for the Physics and Mathematics of
the Universe (IPMU) / University of Tokyo, the Korean Participation
Group, Lawrence Berkeley National Laboratory, Leibniz Institut f\"ur
Astrophysik Potsdam (AIP), Max-Planck-Institut f\"ur Astronomie (MPIA
Heidelberg), Max-Planck-Institut f\"ur Astrophysik (MPA Garching),
Max-Planck-Institut f\"ur Extraterrestrische Physik (MPE), National
Astronomical Observatories of China, New Mexico State University, New
York University, University of Notre Dame, Observat\'ario Nacional /
MCTI, The Ohio State University, Pennsylvania State University,
Shanghai Astronomical Observatory, United Kingdom Participation Group,
Universidad Nacional Aut\'onoma de M\'exico, University of Arizona,
University of Colorado Boulder, University of Oxford, University of
Portsmouth, University of Utah, University of Virginia, University of
Washington, University of Wisconsin, Vanderbilt University, and Yale
University.

The CosmoSim database used in this paper is a service by the
Leibniz-Institute for Astrophysics Potsdam (AIP).  The MultiDark
database was developed in cooperation with the Spanish MultiDark
Consolider Project CSD2009-00064.

The authors gratefully acknowledge the Gauss Centre for Supercomputing
e.V. (www.gauss-centre.eu) and the Partnership for Advanced
Supercomputing in Europe (PRACE, www.prace-ri.eu) for funding the
MultiDark simulation project by providing computing time on the GCS
Supercomputer SuperMUC at Leibniz Supercomputing Centre (LRZ,
www.lrz.de).

The Bolshoi simulations have been performed within the Bolshoi project
of the University of California High-Performance AstroComputing Center
(UC-HiPACC) and were run at the NASA Ames Research Center.

\section{Data Availability}
The data underlying this article are available in
https://skyserver.sdss.org/casjobs/. The datasets were derived from
sources in the public domain: https://www.sdss.org/

\bsp	
\label{lastpage}
\end{document}